\documentclass[%
 reprint,
superscriptaddress,
 amsmath,amssymb,
 aps,
showkeys,
]{revtex4-2}

\usepackage{graphicx}
\usepackage{dcolumn}
\usepackage{bm}
\usepackage{hyperref}
\usepackage{xspace}
\usepackage{soul}

\usepackage{xcolor}

\begin{document}

\preprint{APS/123-QED}

\title{Probing molecular-motor forces driving \textit{Chlamydomonas} gliding motility by holographic microscopy}

\author{Nicolas Fares}
\email{Contact author: Nicolas.Fares@ist.ac.at}
\altaffiliation[Now at ]{Institute of Science and Technology Austria, Klosterneuburg, Austria}
\author{Elorri Garcia}%
\altaffiliation[Now at ]{CBMN, UMR 5248, CNRS, Bordeaux INP, University of Bordeaux, Pessac, France}
\author{Ahmad Badr}%
\author{Yacine Amarouchene}%
\affiliation{%
Univ. Bordeaux, CNRS, LOMA, UMR 5798, F-33400, Talence, France.
}%
\author{Alexandros A. Fragkopoulos}
\altaffiliation[Now at ]{TreeFrog Therapeutics, 33600, Pessac, France.}
\author{Oliver B\"aumchen}
\affiliation{
Experimental Physics V, University of Bayreuth, Bayreuth, D-95447, Germany.
}%
\author{Thomas Salez}
\email{Contact author: thomas.salez@cnrs.fr}
\altaffiliation[Also at ]{Mechanics Department, Ecole Polytechnique, Institut Polytechnique de Paris, 91128 Palaiseau, France}
\author{Antoine Allard}%
\email{Contact author: antoine.allard@u-bordeaux.fr}
\affiliation{%
Univ. Bordeaux, CNRS, LOMA, UMR 5798, F-33400, Talence, France.
}%

\date{\today}

\begin{abstract}
\indent \textbf{Abstract.} The unicellular microalga \textit{Chlamydomonas reinhardtii} is a widely used model microswimmer for biophysicists. 
However, the interest around \textit{C. reinhardtii} goes beyond its swimming capabilities. In fact, light can drastically alter its behavior: under blue illumination, the cell attaches to a nearby surface and intermittently glides on it.
Such a gliding motility is powered by molecular-motor proteins operating in the cell's cilia, and the related machinery has established the cell as a prime reference for the study of intraciliary-transport mechanisms.
In the present work, we track label-free cells using in-line holographic microscopy, which allows us to characterize their gliding motility without requiring fluorescent labeling. 
After benchmarking the experimental technique on swimming cells, we show that, while gliding, the cells move mostly along the surface, and exhibit anomalous-diffusive features, including Lorentzian-like distributions of displacements.
Furthermore, gliding trajectories, by being intermittent, are relevant candidates to determine forces at the molecular-motor scale.
We report a gliding threshold of about 20~pN, which is compatible with the activity of single molecular motors, and on the same order of magnitude as the forces generated during intraciliary cargo transport.
\end{abstract}


\maketitle

Numerous cellular functions, from embryogenesis~\cite{deneke2019self} or cytokinesis~\cite{sedzinski2011polar} to intracellular trafficking~\cite{fletcher2004introduction} and signaling or sensing~\cite{singla2006primary,wang2024primary}, inherently rely on the ability to transport molecules and regulate local stresses. 
Inside the cell, such tasks are often mediated by molecular-motor proteins \textit{walking} along cytoskeletal filaments in the cytoplasm, including those that form the core of organelles like cilia and flagella. These organelles are appendages made of a microtubule-based scaffold, called the axoneme and which consists of parallel microtubule doublets that serve as bidirectional tracks for molecular motors~\cite{gibbons1981cilia,lindemann2024mechanics,stepanek2016microtubule}. 
As \textit{kinesin} and \textit{dynein} motors carry protein complexes along the microtubules, the aptly named intraflagellar-transport (IFT) machinery governs the assembly, maintenance and functions of cilia and flagella ~\cite{rosenbaum2002intraflagellar,huangfu2003hedgehog,wang2006intraflagellar}. 
For instance, cilia and flagella act as sensory and motile organelles~\cite{singla2006primary,scholey2006intraflagellar,christensen2007sensory}.
Consequently, the multifaceted roles of IFT motivated considerable research efforts~\cite{rosenbaum2002intraflagellar,scholey2003intraflagellar,lacey2025intraflagellar} and still triggers modern investigations~\cite{stepanek2016microtubule,lacey2024extensive,zhang2021direct,lewis2024contribution,sun2025intraflagellar,lettermann2024geometrical,bondoc2025functional} aiming at unveiling the mechanisms at play.

The basic mechanisms behind IFT were first uncovered and addressed through the unicellular microalga \textit{Chlamydomonas reinhardtii}~\cite{kozminski1993motility,kozminski1995chlamydomonas,silflow2001assembly}. This cell consists of a spheroidal eukaryotic body, about 10 micrometers in diameter, bearing two cilia whose coordinated beating propels the cell in a breaststroke-like manner~\cite{harris2009chlamydomonas,polin2009chlamydomonas,goldstein2015green,drescher2010direct,souzy2022microbial}.
However, the cell does not simply swim in liquids, potentially under flow~\cite{lagoin_enhanced_2025}, but responds to light stimuli, and the latter can even drastically alter the cell's dynamics. Notably, under blue illumination, the cell attaches to a nearby surface~\cite{kreis2018adhesion} and intermittently glides on it~\cite{till2022motility,jeanneret2016brief}. 
This gliding motility is not driven by ciliary beating~\cite{bloodgood1981flagella} but powered by the aforementioned IFT machinery~\cite{kozminski1993motility,shih2013intraflagellar}.
In more detail, molecular motors are coupled to IFT complexes, which in turn interact with flagellar membrane glycoproteins that mediate adhesion to the substrate~\cite{hoepfner2025unwrapping, shih2013intraflagellar}. These components together form an IFT train~\cite{jeanneret2016brief}. Upon substrate adhesion, IFT trains can generate forces along the axoneme, which are transmitted through the flagellar membrane to the substrate and ultimately to the cell body, thus driving gliding motility.
The intermittent nature of IFT, mixing periods of active linear transport and periods of passive fluctuations~\cite{prevo2017intraflagellar}, was evidenced by correlating the cell body's motion to displacements of IFT trains~\cite{shih2013intraflagellar}, which are triggered by transient calcium elevations within the cilia~\cite{bloodgood1990calcium,collingridge2013compartmentalized,fort2021ca2+}. 
This intermittent nature of IFT was later confirmed from a dynamical perspective at the cell's scale, by focusing on the gliding motility of populations of cells~\cite{till2022motility}. 
Furthermore, the dynamical behavior of adhered cells, powered by IFT trains pulling through ciliary adhesion agents, offers a unique alternative handle on molecular-scale force generation.
Access to gliding trajectories with nanometric precision could reveal the binding and translational forces (\textit{i.e.} normal and transverse to a cilium) that govern motor activity. 

While previous work established the main biological mechanisms of gliding, the relationship between such an intermittent dynamics and the effective molecular forces remains to be addressed. The present study focuses on the \textit{gliding motility} of \textit{C.~reinhardtii}, defined as the intermittent sequence of \textit{active gliding} and \textit{pausing} phases. 
Using low-intensity in-line holographic microscopy~\cite{lee2007characterizing,lavaud2021stochastic}, we quantify the gliding dynamics with nanometer-scale horizontal and submicrometer vertical scale uncertainty, and a 10-millisecond resolution. We then show that cells display anomalous-diffusion features~\cite{metzler2000random}.
We further demonstrate that gliding motility is ultimately driven by the activity of only a few molecular motors, therefore allowing to determine the minimal forces at play in intraciliary transport.

\section*{\label{sec:tracking}Tracking \textit{C. reinhardtii} cells}

\begin{figure}[ht!]
    \centering
    \includegraphics[width=8.5cm]{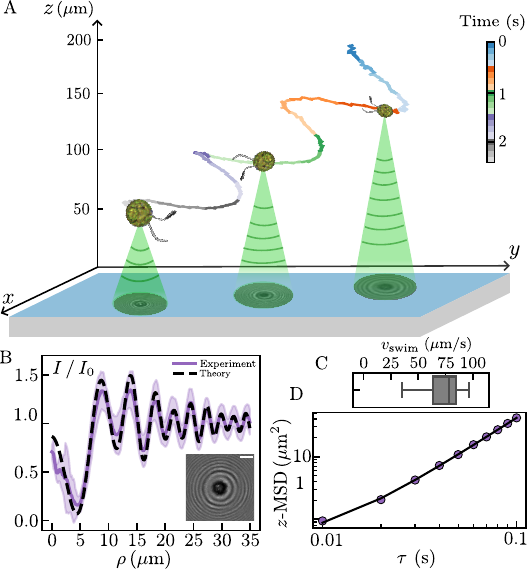}
    \caption{
    Tracking of the unicellular microalga \textit{C. reinhardtii} by holography: benchmark on swimming cells.
    (A) Illustration of a swimming cell in the experimental setup (see Materials and methods). A plane wave illuminates the cell, which scatters light. Holograms, \textit{i.e.} interference patterns between incident and scattered waves, are captured over time and give access to the distance $z$ between the cell and the focal plane of the objective. From the lowest to highest position: $z = 58.0 \, \mathrm{\mu m}$, $z = 91.4 \, \mathrm{\mu m}$, and $z = 151.0 \, \mathrm{\mu m}$ ($x,y$ not to scale).
    (B) Radial intensity profile $I$ (normalized by the incident intensity $I_0$) as a function of the distance $\rho$ from the center of the hologram displayed in inset (scale bar: $10 \, \mathrm{\mu m}$). The purple solid line depicts the experimental profile (orthoradial average), with the light-shaded purple area representing the related error (orthoradial standard deviation). The black dashed line depicts the corresponding theoretical profile, \textit{i.e.} the best fit to the Mie theory~\cite{lavaud2021stochastic} of the two-dimensional image shown in inset, leading to: $a = 3.93 \, \mathrm{\mu m}$ (alga's effective radius), $n_\mathrm{p} = 1.39$ (alga's effective refractive index), and $z = 195.5 \, \mathrm{\mu m}$ (distance between the cell's center and the focal plane of the objective). 
    (C) Box plot of measured total velocities $v_\mathrm{swim} = \sqrt{v_x^2+v_y^2+v_z^2}$. 
    (D) Vertical mean squared displacement ($z$-MSD) as a function of the time lag $\tau$ for the swimming cell depicted in panel~A. The disks correspond to experimental data, and the black solid line to a second-order polynomial fit. As explained in the text, the constant term gives an estimate of the $z$-measurement uncertainty, which is about $360 \, \mathrm{nm}$. 
    }
    \label{fig:swimming}
\end{figure}

The long-time dynamics of gliding \textit{C. reinhardtii} cells was studied previously, but only through two-dimensional tracking resolved at the sub-micrometric scale~\cite{till2022motility,bondoc2025functional}. 
However, cells may fluctuate on smaller scales and tilt or elevate~\cite{fort2021ca2+}. 
Hence, studying their motility could benefit from more-precise microscopy tools able to track microorganisms in three dimensions over a broad time range.
High-precision three-dimensional tracking often relies on techniques that require pre-calibration of the particle's $z$-position, which has proven effective for a single particle~\cite{gosse2002magnetic,lionnet2012single}. To overcome this calibration requirement, alternative approaches have been developed, including interferometry and stereoscopy~\cite{rieu2021parallel}, Lagrangian tracking~\cite{darnige2017lagrangian}, and holography~\cite{barkley2019holographic,lavaud2021stochastic,wang2014using}, all maintaining a sub-micrometric precision when applied to microswimmers.
Furthermore, holography provides information on the particle's orientation and size, consequently enabling novel views on the dynamics of \textit{e.g.} elongated silica rods and Janus particles~\cite{wang2014using}, microdroplets~\cite{fares2024observation}, or swimming bacteria~\cite{wang2016tracking}.

Here, the motion of cells is recorded using Mie holography~\cite{lavaud2021stochastic}, as schematized in Fig.~\ref{fig:swimming}A (green patterns), and detailed in Materials and Methods. 
Briefly, one records the interference patterns, called holograms, between the incident beam (plane waves) and the light scattered by an individual cell. 
Since the scattered field can be computed exactly using Mie theory~\cite{hergert2012mie}, the particle's 3D position $(x,y,z)$, its radius $a$, and its refractive index $n_\mathrm{p}$ are obtained by fitting the experimental hologram, and the resulting theoretical profile is shown in Fig.~\ref{fig:swimming}B. 
We use swimming cells as a benchmark of the technique. The trajectory shown in panel~A follows a helical path, with a time period of 0.75~s and a pitch of $63~\mathrm{\mu m}$, which are close to the expected values for the swimming of \textit{C. reinhardtii}~\cite{leptos2023phototaxis}. 
In addition, the swimming velocities displayed in Fig.~\ref{fig:swimming}C fall in the expected ballpark~\cite{harris2009chlamydomonas,jeanneret2016brief}. 
Therefore, we show that Mie holography enables to accurately capture cell dynamics.

It must be stressed out that Mie holography is well-suited for spherical homogeneous particles. However, cells are spheroidal and inhomogeneous, implying that their tracking is based on the consideration of effective homogeneous spheres characterized by radii $a$ and refractive indices $n_\mathrm{p}$. 
Yet, this \textit{spherical-chlamy} approximation provides a reasonable description of the measured holograms. Indeed, the experimental holograms and the theoretical Mie fits fairly match, as shown in Fig.~\ref{fig:swimming}B, with a radius $a= 3.93 \pm 0.05 \, \mathrm{\mu m}$ and a refractive index $n_{\rm p}=1.39 \pm 0.02$ consistent with reported values~\cite{catalan2025preparation,lee2013spectral,ueki2016eyespot}.

Nevertheless, the approximation impacts the spatial resolution.
Regarding in-plane positions $(x,y)$, the fitting of several fringes leads to an uncertainty of about 5~nm. 
Besides, one can estimate the uncertainty on the $z$-measurement through the mean squared displacement (MSD) in the $z$ direction, and more precisely, through its saturation at short lag time $\tau$.
As displayed in Fig.~\ref{fig:swimming}D and explained in the Materials and Methods, the resulting $z$-uncertainty is on the order of 360~nm.
 
Having benchmarked our ability to resolve the dynamics of freely swimming cells, we now exploit the same approach to probe the qualitatively different motion of gliding cells.

\section*{\label{sec:introgliding}Gliding motility}

\begin{figure}[ht!]
    \centering
    \includegraphics[width=8.5cm]{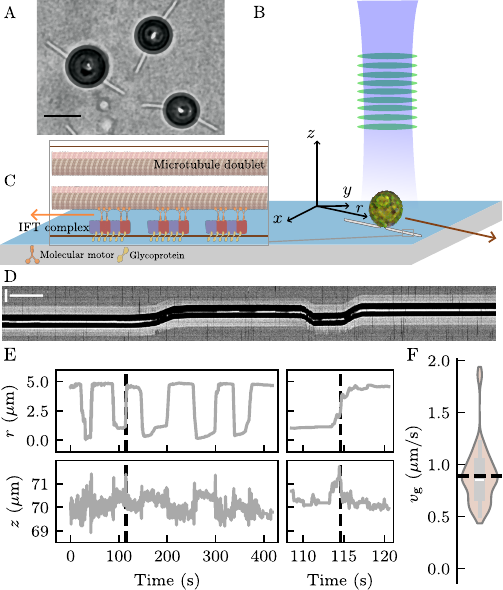}
    \caption{Gliding motility of the unicellular microalga \textit{C. reinhardtii}. 
    (A) Bright-field (Olympus inverted microscope, 20x objective from Olympus) picture showing cells in the gliding configuration (top view, scale bar: $10 \, \mathrm{\mu m}$). 
    (B) Illustration of a gliding cell within the experimental setup. The blue LED illumination favors gliding instead of swimming~\cite{kreis2018adhesion}. The green plane wave is used for holographic tracking. 
    (C) Schematic (not to scale) of the inside of a cell's cilium to highlight the mechanism driving gliding motility (see text).
    (D) Kymograph in the direction of a gliding cell's cilia obtained from a sequence of bright-field images as in panel~A.
    The black (resp. white) areas correspond to the cell's body (resp. cilia).
    Scales: $10 \, \mathrm{\mu m}$ for distances (vertical) and $20 \, \mathrm{s}$ for the time (horizontal). 
    (E) Trajectory of a gliding cell. The top line corresponds to the in-plane $r$-coordinate, with $r = \sqrt{(x - x(0))^2 + (y-y(0))^2}$, and the bottom line to the $z$-coordinate. The right-hand-side column of the panel corresponds to a zoom on one gliding event, indicated by the black dashed lines. 
    (F) Violin plot showing the distribution of in-plane gliding speeds $v_\mathrm{g}$ of cells. The dashed line represents the average velocity, which is about $0.9 \, \mathrm{\mu m / s}$ (measured over 343 gliding events, 23 cells). 
    }
    \label{fig:intro}
\end{figure}

A \textit{C. reinhardtii} cell responds to a blue illumination by adhering to surfaces~\cite{kreis2018adhesion}. When adhering to a surface, the cell adopts a specific configuration, which is displayed in Fig.~\ref{fig:intro}A,B (top and side views). 
The two cilia of the cell extend approximately 180\textdegree~apart from each other, and bind to the surface thanks to glycoproteins located on their membranes~\cite{jeanneret2016brief,hoepfner2025unwrapping}. Once attached, the cell nonetheless glides along the surface~\cite{till2022motility,jeanneret2016brief}. 
As schematized in Fig.~\ref{fig:intro}C, this motion arises from intraflagellar-transport (IFT) trains~\cite{shih2013intraflagellar}: IFT complexes connect the surface-bound glycoprotein layer of cilia to molecular motors, which walk along the outer microtubule doublets of the axoneme, from the ciliary tip toward the cell body (orange arrow, panel~C). Because the glycoproteins remain anchored to the substrate, the motor-driven motion of the IFT trains pulls the cell body forward (brown arrow, panel~B).
Consequently, cell motion is driven by the intermittent engagement of a small subset of IFT trains, likely localized on microtubule doublets in the vicinity of the substrate, overcoming cilia-surface adhesion. Since the effective engagement of these IFT trains fluctuates over time, the cell intermittently switches between (i) back-and-forth active gliding displacements along a preferred in-plane direction, at a typical speed of approximately $1\, \mathrm{\mu m / s}$~\cite{till2022motility}, when the number of trains is greater on one cilium, and (ii) pausing periods, when the number of trains on each cilium is equivalent.
The combination of (i) and (ii) defines gliding motility. 
Conversely, a cell is considered as \textit{resting} if its position has not significantly changed within 500 seconds.
In addition, note that cells actually glide -- rather than ``toggle'' -- on the bottom surface of the chamber. This is confirmed by observing simultaneously cells' bodies and cilia, and by correlating their motion, as highlighted on the kymograph displayed in Fig.~\ref{fig:intro}D. 
The black area, \textit{i.e.} the cell's body, follows the exact same trend as the white areas, \textit{i.e.} the cell's cilia. 

Strikingly, rather than exploring the whole surface to which they adhere, cells often stochastically oscillate between two distinct in-plane positions (Fig.~\ref{fig:intro}E, upper panels). 
This restricted two-position behavior is unexpected, but offers a convenient and well-defined framework to unveil general gliding features. Indeed, motions between these two positions directly reflect the activity of the cilia that remain attached to the substrate. Importantly, the gliding speeds measured during these transitions (Fig.~\ref{fig:intro}F) fall in the expected velocity range of about $1\,\mathrm{\mu m /s}$, reported for standard gliding~\cite{till2022motility}. This agreement indicates that the underlying dynamics remains governed by the same molecular processes.

Let us now focus on the $z$-coordinate of a gliding cell, as shown in Fig.~\ref{fig:intro}E. An elevation of the cell while a cilium re-positions on the surface was suggested in Ref.~\cite{fort2021ca2+}. 
Here, zooming on gliding events (right panels, Fig.~\ref{fig:intro}E) seems to indicate a correlation between the onset of in-plane gliding displacements with an out-of-plane elevation of the cell. 
However, the magnitude of the observed elevations is comparable to the $z$-uncertainty of $0.4 \, \mathrm{\mu m}$, estimated earlier in the manuscript. 
Also, the aforementioned \textit{spherical-chlamy} approximation may hide tilting of the cell body.
These limitations prevent us from drawing definitive conclusions regarding the out-of-plane dynamics, but point to avenues for additional investigation. 
Nonetheless, our three-dimensional tracking presented here confirms that gliding is predominantly an in-plane motion.

\section*{\label{sec:gliding}Anomalous diffusive behavior}

\begin{figure}[ht]
    \centering
    \includegraphics[width=8.5cm]{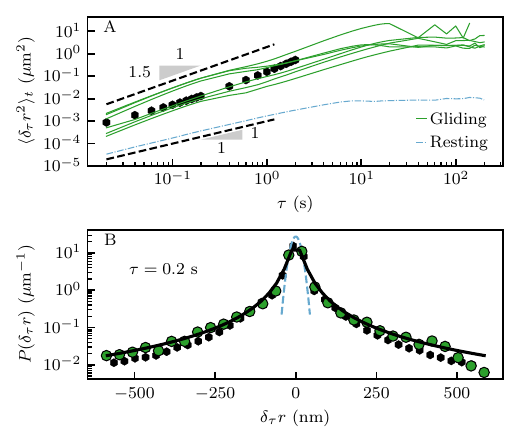}
    \caption{Anomalous, super-diffusive-like behavior of gliding cells. 
    (A) In-plane mean squared displacements (MSDs) $\langle \delta_\tau r^2 \rangle_t$ as functions of the lag time $\tau$. Colored lines correspond to experimental MSDs from individual trajectories, either (green) when the cells exhibit the intermittent gliding motility (6 cells) displayed in Fig.~\ref{fig:intro}E or (blue) for our longest \textit{resting} cell (more than 500~seconds). The dashed lines are guides to the eye. 
    (B) Probability distribution function (PDF) $P(\delta_\tau r)$ of observing in-plane displacements $\delta_\tau r$ over a time lag $\tau$. The green disks correspond to intermittent trajectories, the black solid (resp. blue dashed) line is a Lorentzian (resp. Gaussian) guide to the eye. 
    In both panels, the black hexagons correspond to the numerical simulations combining Brownian motion and Lévy walks as described in the Materials and Methods. 
    }
    \label{fig:gliding_dynamics}
\end{figure}

Dynamical features of gliding motility are investigated hereafter, focusing on the in-plane motion of cells, as shown in~\ref{fig:intro}E (upper panels). 
The observed two-state dynamics is further investigated through the mean-squared displacements (MSDs), $\langle \delta_\tau r^2 \rangle_t$, and the distributions $P(\delta_\tau r)$ of in-plane displacements $\delta_\tau r = r(t+\tau) - r(t)$ over a lag time $\tau$ at time $t$, with $\langle\cdot\rangle_t$ denoting the average over $t$. Those observables, which are displayed in Fig.~\ref{fig:gliding_dynamics}, highlight the anomalous super-diffusive-like behavior~\cite{metzler2000random,mendez2014stochastic,munoz2025quantitative} of cells, consistently revealing that intermittently gliding cells differ from pure Brownian entities.
More specifically, as evidenced by the green lines in~panel~A, the MSDs of cells exhibiting a two-position gliding motility scale as $\tau^\beta$, with $\beta\simeq 1.3 -1.5$, at short lag times, rather than the linear scaling expected for Brownian diffusion~\cite{bian2016111}.
This anomalous scaling is mirrored in the displacement statistics (see panel~B), where the experimental distribution $P(\delta_\tau r)$ (green dots) deviates from a purely-Brownian Gaussian statistics (black dashed line), with tails better captured by a Lorentzian law (black solid line). 
Besides, as shown in Supplementary Information (SI, see Fig.~S2), this Lorentzian-like shape is observed for several time lags $\tau$, and the widths of the Lorentzian-like distributions are $\tau$-dependent and rescalable. 

Super-diffusive-like dynamics are often associated with enhanced search strategies, as encountered for intracellular cargoes that alternate between passive diffusion and active motor-driven transport~\cite{prevo2017intraflagellar,sittewelle2024passive}. 
Interestingly, a similar interplay is observed for gliding cells -- here, between pausing Brownian phases and active gliding displacements.
To corroborate this statement, let us examine one cell that remains at a fixed location for more than 500~seconds, \textit{i.e.} which is resting. As evidenced by the blue curve in~panel~A, the MSD of this resting cell grows linearly with time (at short time), consistent with the diffusive regime of a trapped purely-Brownian particle. Fitting this initial regime of the MSD by $4D_\mathrm{rest}\tau$ leads to $D_\mathrm{rest} = 8.8 \cdot 10^{-3} \, \mathrm{\mu m^2/s}$. 
The magnitude of $D_\mathrm{rest}$ enables us to estimate an effective friction coefficient $\gamma_\mathrm{rest}=k_{\rm B}T/D_{\rm rest}\simeq 4.5 \cdot 10^{-7} \, \mathrm{kg/s}$, larger than the one expected for a freely diffusing spherical particle of radius $a= 4 \, \mathrm{\mu m}$ ($\gamma\simeq 8 \cdot 10^{-8} \, \mathrm{kg/s}$). Therefore, this result is consistent with enhanced dissipation due to cilia-surface interaction, and $D_\mathrm{rest}$ represents an effective diffusion coefficient of cells confined through surface adhesion.

Motivated by the aforementioned intermittent nature of the gliding motility, we mimic the cells' dynamical behavior (displayed in Fig.~\ref{fig:gliding_dynamics}) through the simulation of trajectories of particles that stochastically switch between two random processes: a purely-Brownian process and an anomalous-like one based on L\'evy walks~\cite{metzler2000random} (see Materials and Methods for details). 
The resulting MSD and distribution of displacements are shown as black hexagons in Fig.~\ref{fig:gliding_dynamics}, and fairly capture the experimental data.
Especially, this model is promising since it leads to distributions of displacements that capture the more-characteristic experimental \textit{extremely-heavy-tailed} distributions.
In fact, other anomalous-diffusion models, such as intermittent fractional Brownian motion~\cite{balcerek2023modelling}, do not feature such a characteristic functional shape, even if they would provide a closer representation of the MSD trend. 
Finally and remarkably, the simulations focus on the unbounded gliding dynamics, while the experimentally observed oscillations between two well-defined positions (Fig.~\ref{fig:intro}E) manifest only through the saturating MSDs at long times (Fig.~\ref{fig:gliding_dynamics}A).
Besides, the same trends in MSDs and displacement distributions are retrieved in comparative standard-microscopy experiments on a population of gliding cells, as shown in SI. 
This suggests that the super-diffusive-like behavior of the cells is a general feature of their gliding motility. 

All in all, this model approach provides a phenomenological description of gliding dynamics rather than a theory linked to IFT activity, but still results from microscopic interpretations at the molecular-motor scale. 
First, the intermittent nature of the gliding dynamics is described \textit{e.g.} in Ref.~\cite{prevo2017intraflagellar}, and is directly related to the intermittent nature of the motion of molecular motors and of their attachment to surfaces or equivalently to cargo molecules, as shown \textit{e.g.} in Ref.~\cite{shih2013intraflagellar}. Second, a L\'evy-like dynamics, as the one observed experimentally in~Fig.~\ref{fig:gliding_dynamics}B, has been reported in the symmetric case of molecular-motor-driven transport of cargo molecules in~\cite{fedotov2018memory}. In addition, heavy-tailed distributions of displacements have been predicted in Ref.~\cite{klumpp2005cooperative} in a similar context. 

\section*{\label{sec:discussion}Force estimation}

Going beyond the observables displayed in~Fig.~\ref{fig:gliding_dynamics}, let us now estimate the force required to initiate gliding, and thereby the minimal number of IFT trains that effectively transmit force to the substrate (Fig.~\ref{fig:intro}C). To access this weak-activity regime, we take advantage of the stationary state reached under prolonged blue illumination, during which gliding activity decays to a level consistent with the competition of only a few IFT trains~\cite{till2022motility}. In this regime, cells exhibit two-position intermittent trajectories (Fig.~\ref{fig:intro}E), which provide a clear signature of the smallest force capable of displacing them.

\begin{figure}[ht!]
    \centering
    \includegraphics[width=8.5cm]{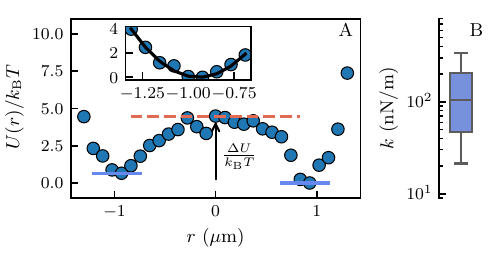}
    \caption{The cells exhibit an effective double-well positional landscape. 
    (A) Potential $U$, normalized by the thermal energy $k_\mathrm{B}T$, as a function of the in-plane position $r$. Blue disks correspond to experimental data. Colored lines are guides to the eye. The inset shows a zoom on the left-hand-side well. The black line corresponds to the best fit to $k(r-r_\mathrm{well})^2/2$, leading to $k= 234 \, \mathrm{nN/m}$ and $r_\mathrm{well} = - 1.0 \, \mathrm{\mu m}$. 
    (B) Box plot showing spring constants $k$ obtained as in panel~A, for 6 different wells (3 cells).
    } 
    \label{fig:potential}
\end{figure}

First, such an intermittent gliding of cells can naturally be treated as a motion within an effective double-well potential (Fig.~\ref{fig:potential}A). The normalized effective potential $U(r)$, reconstructed from the in-plane position $r$ (see Materials and Methods), clearly displays two stable wells. Each well is locally and accurately approximated by a harmonic potential (inset, Fig.~\ref{fig:potential}A), whose curvature leads to spring constants on the order of $0.1 \, \mathrm{\mu N/m}$ (Fig.~\ref{fig:potential}B). 
Furthermore, the minimum effective resting force required to cross the barrier and initiate a transition between the wells, can be estimated from the ratio $\Delta U / \Delta r$, where $\Delta U$ is the barrier height separating the wells and $\Delta r$ their spatial separation. 
This leads to forces of a few tens of femtonewtons, which can be interpreted as a lower bound on the force required to trigger gliding events. Interestingly, this force is several orders of magnitude below the force generated by a single IFT train~\cite{shih2013intraflagellar}, consistent with the idea that gliding is initiated by rare and transient events overcoming weak cilia-surface interaction.

\begin{figure}[ht!]
    \centering
    \includegraphics[width=8.5cm]{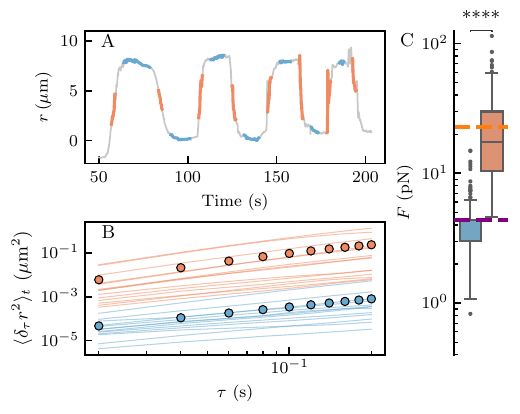}
    \caption{Forces acting on single cells. 
    (A) In-plane coordinate $r$ as a function of time. The colors highlight periods of motion (red) and periods of pausing at a given location (blue). 
    (B) In-plane MSDs $\langle \delta_\tau r ^2\rangle_t$ as functions of the time lag $\tau$, in both aforementioned regimes. The MSDs correspond to one cell. The lines correspond to single intervals, either gliding or pausing. The disks correspond to the average of the lines, in both of the aforementioned regimes.
    (C) Box plot showing the in-plane forces $F$ measured according to Eq.~(\ref{eq:force}), with $\tau = 20 \, \mathrm{ms}$, in both of the aforementioned regimes (7 cells, leading to 153 gliding events). The orange (resp. purple) dashed line corresponds to the mean force measured when the cells glide (resp. pause at a given location). The stars indicate the result of an independent Welch's t-test with a $p$-value of $10^{-9} \ll 0.001$, thus showing that the two means differ significantly (without assuming equal variances between the distributions).
    }
    \label{fig:forces}
\end{figure}

Second, in order to investigate the forces at play in greater detail, periods when cells actively glide on the surface are separated from periods when the cells pause at a given location, as illustrated in Fig.~\ref{fig:forces}A. 
Note that, in the latter figure, the orange colors refer to gliding events and the blue colors symbolize pauses between gliding events.
Then, in-plane MSDs $\langle \delta_\tau r ^2 \rangle_t$ are computed in both regimes and displayed in Fig.~\ref{fig:forces}B. As expected, the MSDs corresponding to the gliding regime are substantially higher than the MSDs corresponding to the pausing regime, \textit{i.e.} the cells are substantially more motile while gliding. 
Besides, the time exponent of the MSDs is greater than one during gliding periods, but equal to one during pausing periods. 

How to relate such an intermittent two-position gliding dynamics to the forces exerted on the cells is yet to be understood. One attempt to do so is discussed hereafter. Let us assume that a cell body experiences an effective drag characterized \textit{via} the drag coefficient $\gamma_\mathrm{pause}$. Furthermore, for the cell body to glide, IFT trains must be in excess on one of the cell's cilia. Then, let us introduce the number $n$ of IFT trains in excess on one of the cell's cilia. Assuming that each IFT train exerts a force $f$ on the cell body and neglecting inertia (which is 3 orders of magnitude smaller than the drag term), Newton's second law reads: 
\begin{equation}
    \label{eq:pfd_gliding}
    0 = - \gamma_\mathrm{pause} \frac{\delta_\tau r}{\tau} + nf \, ,
\end{equation}
in which the drag coefficient $\gamma_\mathrm{pause} = k_\mathrm{B}T / D_\mathrm{pause}$ is measured by fitting the pause-period ensemble-averaged MSDs displayed in Fig.~\ref{fig:forces}B with $4D_\mathrm{pause}\tau$. 
We measure the drag coefficient using pausing phases, assuming that the fluctuations can be described by an equilibrium Einstein relation. The corresponding drag is on the order of $\gamma_\mathrm{pause}\simeq8\cdot 10^{-6} \, \mathrm{N/m\cdot s}$. Note that this value is larger than $\gamma_\mathrm{rest}$, indicating that the pausing cell is more active than the resting one, which is reminiscent of a higher molecular activity that enables triggering a gliding movement. The equilibrium value $\gamma_\mathrm{pause}$ is then used in a near-equilibrium linear response of gliding motion, in which the force can be estimated from the observed displacement, following:
\begin{equation}
    \label{eq:force}
    F = \sqrt{\langle (nf)^2\rangle_t} = \gamma_\mathrm{pause} \frac{\sqrt{\langle \delta_\tau r ^2\rangle_t}}{\tau} \, .
\end{equation}
The averaging is performed on the gliding and pause parts of the cell's trajectory separately, and the resulting forces are displayed in Fig.~\ref{fig:forces}C. 
The corresponding measurements in the $z$-direction are displayed in SI for the sake of completeness.
The median forces measured are about 18~pN and 4~pN when the cells are gliding or are pausing, respectively. 
Furthermore, comparing the highest forces acting on cells pausing at a given location (outliers corresponding to the blue box in~Fig.~\ref{fig:forces}C) and the average force acting on gliding cells (orange dashed line in~Fig.~\ref{fig:forces}C) indicates the existence of a force threshold for cells to glide of about 20~pN.
Such a threshold is comparable to the force expected from a single IFT train.
Indeed, an IFT train generates approximately 25~pN according to the direct optical-trapping stall-force measurements performed in Ref.~\cite{shih2013intraflagellar}. 
Hence, the present measurement suggests that gliding motility is ultimately powered by at most one or two IFT trains at a time.
Finally, let us stress that, surprisingly, the forces measured while cells are gliding fall in the same ballpark as the 26-piconewton-like forces generated by beating cilia~\cite{boddeker2020dynamic}.

\section*{\label{sec:ccl}Conclusion}

We use holographic microscopy to investigate the gliding motility of \textit{C. reinhardtii} cells.
Our approach therefore enables mechanical measurements to be extracted from long-term trajectories, using low-intensity illumination, and providing a complement to fluorescence-based direct tracking of IFT trains~\cite{shih2013intraflagellar,fort2021ca2+,stepanek2016microtubule}.
Specifically, three-dimensional holography reveals that gliding remains predominantly in-plane, with heavy-tailed anomalous-like dynamics.
Furthermore, gliding trajectories, by being intermittent, allow one to determine minimal forces at the molecular-motor scale.
Here, a gliding-displacement threshold of about 20~pN is observed, which is consistent with gliding being triggered by an imbalance of one or two IFT trains between the cells' two cilia.
Interestingly, repeated oscillations between two given positions, instead of unbounded random walks, are observed. The aforementioned anomalous-like behaviors do not stem from this two-position dynamics, but the latter dynamics and its origin are yet to be understood. 
One potential explanation resides in the frequency and amplitude of calcium elevations in the cilia, which are known to regulate the activity of IFT trains~\cite{bloodgood1990calcium,collingridge2013compartmentalized} and the interactions between IFT complexes and the surface-binding proteins~\cite{fort2021ca2+}.

\section*{Materials and Methods}

\subsection*{Mie holography}

The microscopy techniques, the \textit{C. reinhardtii} strains used, the cultivation procedure, and the experimental protocol are described hereafter. 
First and foremost, cells are tracked by in-line holography~\cite{lee2007characterizing,lavaud2021stochastic}. 
As illustrated in Fig.~\ref{fig:swimming}A, a coherent green plane wave (laser CPS532 from Thorlabs, wavelength 532~nm, output power 4.5~mW) illuminates a cell, which scatters part of the incident wave. Then, the scattered wave interferes with the non-scattered part of the incident wave, leading to holograms as shown in Fig.~\ref{fig:swimming}B. These holograms, recorded through an Olympus oil-immersion x100 objective and a camera (ac-1920um from Basler), are fitted to the Lorenz-Mie theory~\cite{hergert2012mie} \textit{via} the open-access Pylorenzmie software suite~\cite{altman2021holographic,lavaud2021stochastic}. Those holograms depend on five parameters: the radius $a$ and refractive index $n_\mathrm{p}$ of the cell, and its three-dimensional position $(x,y,z)$. However, we stress out here that the cell is spheroidal rather than spherical, and inhomogeneous. 
Hence, its tracking relies on the consideration of an effective sphere in the Mie theory, which applies to spherical homogeneous scatterers. 
Such an approximation was previously made for instance to track dimers by holography~\cite{altman2021holographic}.
Yet, the shape of an experimental hologram corresponding to the cell is fairly captured by the theory for spherical scatterers, as shown in Fig.~\ref{fig:swimming}B, with values of radius and refractive index that are in fair agreement with the literature~\cite{catalan2025preparation,lee2013spectral} as detailed in SI. Practically, holograms are captured far away from the cell in the $z$-direction to consolidate the spherical approximation.

\subsection*{Estimate of the $z$-uncertainty}

The uncertainty on the $z$-measurement is estimated on a swimming cell, through the mean squared displacement (MSD) in the $z$ direction.
Specifically, let us assume that $z = z_\mathrm{r} \pm u_z$, with $u_z$ denoting the magnitude of the uncertainty on the measured position $z$, and variables with the subscript r denoting the real values (as opposed to the measured values with no subscript). Displacements $\delta_\tau z$ over a time lag $\tau$ at time $t$ read $\delta_\tau z = \delta_\tau z_r \pm 2u_z$. Then, squaring this equation and averaging over $t$ leads to $z\text{-MSD} = \langle \delta_\tau z^2 \rangle_t = z\text{-MSD}_\mathrm{r} + (2u_z)^2$, by assuming position-noise independency. Hence, a short-time saturation of the $z$-MSD reflects measurement uncertainty. Accordingly, as shown in Fig.~\ref{fig:swimming}D, we fit the measured $z$-MSD of a swimming cell to a quadratic polynomial -- that accounts for a diffusive regime and a ballistic one -- and thus obtain an estimate of the uncertainty $u_z$ on the order of 360~nm.

\subsection*{Monitoring light conditions}

In order to monitor the cells' motility, a blue LED (470 nm, purchased from Thorlabs\copyright, reference: M470) and a red laser (670~nm, output power 4.25~mW, purchased from Lasermax, Inc.) illuminate the sample obliquely. The blue light intensity is $10^{20} \, \mathrm{ph \cdot m^{-2}\cdot s^{-1}}$ but the actual intensity illuminating the suspension remains hard to quantify precisely as the light is directed to the suspension at an angle.

\subsection*{Strains and cultivation}

The main \textit{C. reinhardtii} strain employed is the CC-409 (purchased from the culture collection \textit{Chlamydomonas Resource Center} CC, which is equivalent to the strain SAG11-32b from \textit{Sammlung von Algenkulturen Göttingen} SAG). 
After a period of culture on solid agar, the cells are cultivated in a home-made Tris-Acetate-Phosphate (TAP) growth medium, which is adapted from the ``Gorman-Levine'' recipe~\cite{gorman1965cytochrome} and resembles the commercial TAP medium from Gibco\copyright, for at least two weeks in order for the cells to adapt and synchronize. Cultures are grown in an incubator, which is maintained at 20\textdegree C, and in which the cells follow a ``day-night'' cycle, with 14-hours day period at a light intensity of $10 \, \mathrm{\mu mol \cdot m^{-2} \cdot s^{-1}}$ and 10-hours dark period. For the experiments, cultures are grown in 10 mL of TAP medium and diluted every day to a concentration of approximately $10^6$~cells/mL. 
Besides, as mentioned in the text corresponding to Fig.~\ref{fig:gliding_dynamics}
and shown in SI, comparative population-level gliding assays were performed by conventional bright-field microscopy. For these experiments, the strain SAG11-32b was employed and cultivated following the recipes in Ref.~\cite{catalan2025preparation}. Protocols and results are detailed in the SI.

\subsection*{Experimental procedure}

On a day of experiments, a suspension of cells prepared two days before is taken in the morning, filled in a 200-$\mu$m-thick glass chamber, and the chamber is put in the microscope. Then, the chamber is left under red light for at least an hour to let the cells grow back their cilia. Then, an adaptation phase, consisting of two cycles of 2 minute darkness and 2-minute red light, followed by 5 minutes under blue light and 15 minutes under red light, is done. This adaptation-phase protocol is adapted from~\cite{till2022motility} with some modifications. After this adaptation phase, an experiment is run. Between each experiment, cells are left under red light for 20 minutes to eliminate any memory-like effects. No experiment is run after 5~pm to avoid the commitment phase of the cells~\cite{catalan2025preparation}. More importantly, as opposed to the work done in Ref.~\cite{till2022motility}, recordings of gliding trajectories start after at least 5 minutes under blue light. This choice allows to neglect the dependency of the gliding dynamics on the time under blue illumination and to study the cells' dynamics in a low-activity permanent regime~\cite{till2022motility}. In addition, this low-activity regime is associated with less de-attachments of the cells' cilia and thus less reorientations. Consequently, the low-activity regime also selects one-dimensional-like dynamics over two-dimensional dynamics.

\subsection*{Simulations}

Motivated by the intermittent nature of the gliding motility, we phenomenologically and numerically approach the dynamics of gliding \textit{C. reinhardtii} by simulating trajectories of particles that stochastically switch between two random processes: a Brownian process and an anomalous-like one.
On the one hand, including a Brownian dynamic comes from noticing that a cell still fluctuates when it does not actually glide on the surface, and that those fluctuations are Brownian, as discussed earlier in the main text related to Fig.~\ref{fig:gliding_dynamics}.
On the other hand, including an anomalous-like dynamic comes from the fact that the dynamic of gliding cells differs from a purely-Brownian behavior. The Lorentzian-like distribution displayed in Fig.~\ref{fig:gliding_dynamics}B motivates the choice of a L\'evy walk~\cite{metzler2000random} to describe gliding. Indeed, a L\'evy walk consists in a random walk with varying step lengths that are sampled from a L\'evy distribution of parameter $\alpha \in [1, 2]$, and give rise to heavy-tailed distributions of displacements. Note that $\alpha=2$ corresponds to a Brownian process and $\alpha=1$ to \textit{extreme flights}. Here, the Lorentzian distribution motivates the choice of $\alpha = 1$, which should capture the distribution's extremely-heavy tails.  
The others parameters of the L\'evy-walk dynamics are chosen to match the experimental data: 
the diffusion coefficient is $D_\mathrm{rest}$ (defined in the text related to Fig.~\ref{fig:gliding_dynamics}), the velocity of the displacements is $1 \, \mathrm{\mu m/s}$, and the displacements are bounded according to the minimal ($0.1\,\mathrm{\mu m}$ over 1~second) and maximal ($2\,\mathrm{\mu m}$ over 1~second) displacements observed experimentally. 
Finally, the times over which the particle behaves according to a given random process are sampled from a uniform distribution, while ensuring that the particle behaves according to the Brownian dynamics at least half of time. 
In total, 100 simulations of 400-second long trajectories, with an integration time of 0.02 second, are run. The resulting MSD and distribution of displacements are shown as black hexagons in Fig.~\ref{fig:gliding_dynamics}, and fairly captures the experimental data.

\subsection*{Experimental in-plane potential} 

The in-plane potential $U(r)$ in which a cell lies is reconstructed from the cell's $r$-trajectory. First, the probability of presence $P(r)$ of the cell in the $r$-direction is obtained by binning the $r$-space and counting the number of appearances of the cell in each bin. Then, the potential $U(r)$ is computed as $U = - k_\mathrm{B}T \ln{P}$.\\

\section*{Acknowledgment}

The authors thank Gabriel Amselem, Juho Lintuvuori, Yoann de Figueiredo, and Matteo Bassanini for interesting discussions. 
The authors thank the Göttingen Algae Culture Collection (SAG) for providing strain SAG 11-32b. 
The authors acknowledge financial support from the European Union through the European Research Council under EMetBrown (ERC-CoG-101039103) grant. Views and opinions expressed are however those of the authors only and do not necessarily reflect those of the European Union or the European Research Council. Neither the European Union nor the granting authority can be held responsible for them. The authors also acknowledge financial support from the Agence Nationale de la Recherche under EMetBrown (ANR-21-ERCC-0010-01), Softer (ANR21-CE06-0029), Fricolas (ANR-21-CE06-0039), and Babin (ANR-25-CE30-3613-01), as well as from the Interdisciplinary and Exploratory Research program under MISTIC grant at University of Bordeaux, France. They also acknowledge the support from the LIGHT S\&T Graduate Program (PIA3 Investment for the Future Program, ANR-17EURE-0027). Besides, N.F. is the recipient of a PhD grant from the Ecole Normale Supérieure de Lyon and the Réseaux de Recherche Impulsion (RRI) ``Frontiers of Life'' which received financial support from the French government in the framework of the University of Bordeaux’s France 2030 program. Finally, the authors thank the Soft Matter Collaborative Research Unit, Frontier Research Center for Advanced Material and Life Science, Faculty of Advanced Life Science at Hokkaido University, Sapporo, Japan, and the CNRS International Research Network between France and India on ``Hydrodynamics at small scales: from soft matter to bioengineering''.

\bibliography{bibliography}

\end{document}